# Multi-Agent System Interaction in Integrated SCM

**Ritu Sindhu[1], Abdul Wahid[2], G.N.Purohit[3]**

**[1]Ph.D Scholar, Banasthali University and WIT
Gurgaon, India**

**[2]Department of CS, AKG Engineering College
Ghaziabad, India**

**[3] Dean, Banasthali University, India**

## Abstract

Coordination between organizations on strategic, tactical and operation levels leads to more effective and efficient supply chains. Supply chain management is increasing day by day in modern enterprises.. The environment is becoming competitive and many enterprises will find it difficult to survive if they do not make their sourcing, production and distribution more efficient. Multi-agent supply chain management has recognized as an effective methodology for supply chain management. Multi-agent systems (MAS) offer new methods compared to conventional, centrally organized architectures in the scope of supply chain management (SCM). Since necessary data are not available within the whole supply chain, an integrated approach for production planning and control taking into account all the partners involved is not feasible. In this study we show how MAS architecture interacts in the integrated SCM architecture with the help of various intelligent agents to highlight the above problem.

## 1. Introduction

Inter-enterprise collaboration is highly challenged and supply chain management (SCM) is one of such strategies through which individual enterprises form partnerships and achieve inter-enterprise collaboration. A supply chain comprises multiple enterprises to collaboratively provide customers with products or services. SCM is the relationship across the supply chain. Multi-agent system (MAS) has been recognized as a promising technology for the automation of SCM in recent years.

The goal of this paper is to introduce a new supply chain planning, execution and control approach with the help of multi-agent systems that perform both inter- and intra-organizational planning and execution tasks are an important innovation. Supply chains consist of networks formed by co-operating partners that are covering various companies (Original Equipment manufacturers (OEM), suppliers, sub-suppliers, etc.).Due to Bullwhip effect of fluctuations in market, the global optimization of the corresponding business processes offers a vast optimization potential. That hampered by the fact that the companies are not willing to reveal their production data to competitors, unless they are forced to do so. due to which a global optimization is hardly feasible today. That's why MAS perfectly suit these demands for global flexibility, co-operation and local autonomy. The individual projects that are involved in the research activities presented in this paper[1][3][4][6] address these problems and offer services in the range of SCM scheduling. In this paper, a reference model integrating the mentioned MAS is introduced including interfaces and gateways between the systems.

### 1.1 SCM Integrated Architecture

The basic scenario is to introduce the reference model comprises a simplified supply chain of a manufacturer of automobile equipment (see figure 1). Its main value to assembly the parts from different suppliers. In this scenario, a manufacture of automobile components and a supplier of tyres are incorporated. The logistics service providers are necessary to deliver the parts to the producer and to its customers.[2][7].

The complexity of managing supply chains results in many different interdependent tasks such as the planning, execution and controlling of production, routing, logistics and inventory processes. The basic scenario focuses on production processes with the help of integration between multiple MAS. Table 1 gives an overview on the various functionality of each MAS involved

The distributed global planning of supply chain activities achieved with the help of DISPOWEB[3] system that generates an initial plan of orders and suborders concerning prices and time-points of delivery, and after that software agents located at the different supply chain partners carry on negotiations[13]. (In Fig 2) We have shown the interaction in integrated architecture with the help of MAS.

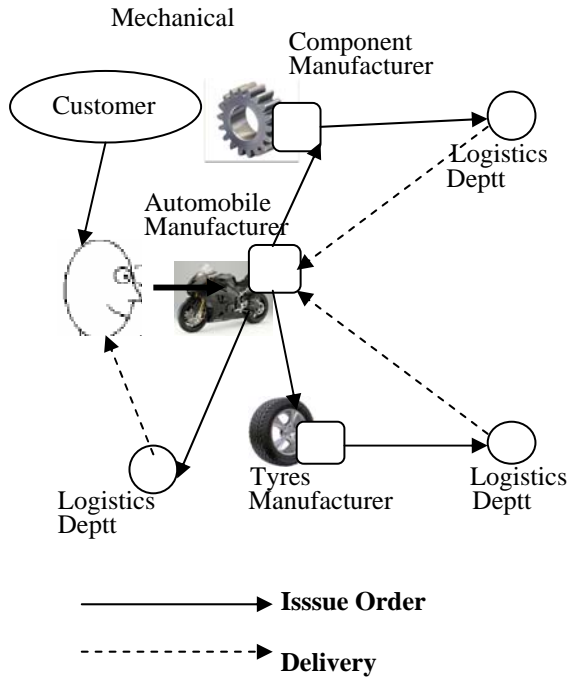

Fig 1. Scenario of SCM

**Table 1. Overview of MAS functionality**

| Main Functionality | Project |
|---|---|
| Negotiations between enterprises | DISPOWEB |
| Integrated process planning and scheduling (with focus on discrete manufacturing) | IntaPS |
| Production planning and controlling (with focus on assembling industries) | KRASH |
| Production planning and controlling (with focus on batch production) | FABMAS |
| Operational tracking of orders including suborders in supply chains | ATT* |
| Analysis of historical tracking information (tracing) | SCC* |

*\* ATT and SCC are conducted in one project*

➢ Firstly supply chain partners optimize the cost and due date of deliveries which are used in each enterprise at intra-level to plan the production of goods in SCM (a).

➢ For production planning three multi agent systems are concerned in which input are required from the DISPOWEB[3] agents for production facilities (b).

➢ Thirdly there are ATT/SCC*[1] project that monitor the orders on every stage of supply chain using a distributed architecture. In case of event that endanger the planned fulfillment. The ATT* system communicates with the partner of enterprise to inform about the endangerment(c)[14].

➢ Fourthly if such type of event occur in an enterprise than the output comes from ATT* system is used to reschedule the process planning to overcome such danger events in future (d).

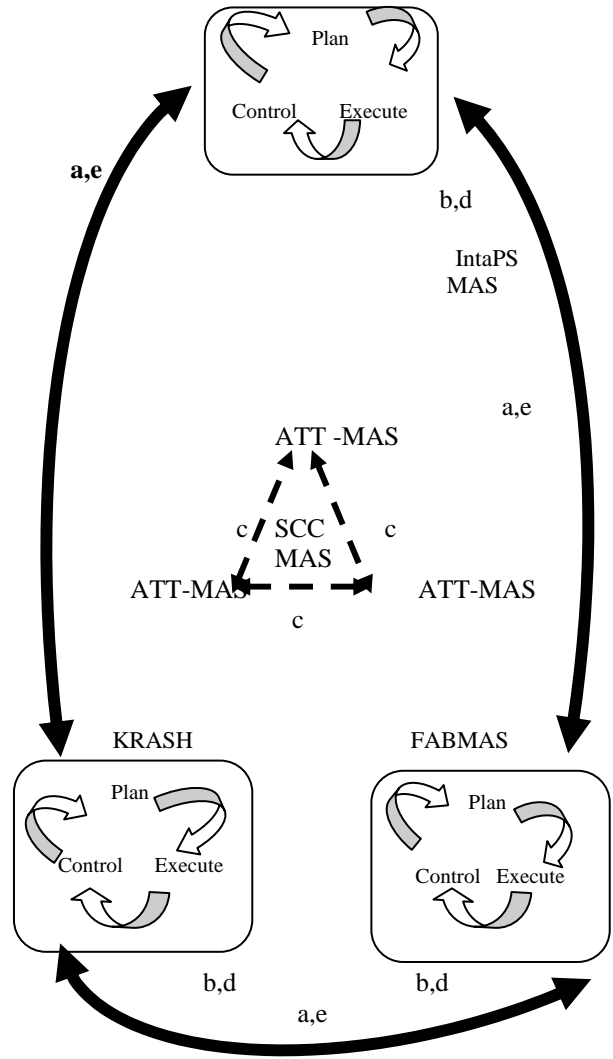

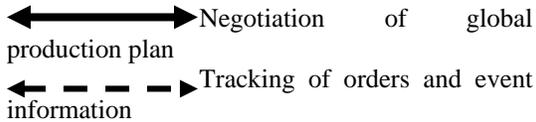

Negotiation of global production plan

Tracking of orders and event information

Fig.2 Architecture of MAS in integrated SCM

➢ In case of such a major events, renegotiation of the contracts at the inter-enterprise level occurs (e). Finally the ATT* system communicate with trusted third party service called SCC* MAS analyze history of orders & their sub-orders due to which we identify the pattern in supply chain orders that leads to the hindrances during fulfillment of planning.

## 2. Database Structure in Integrated SCM

The data structure is a significant component of the presented approach and serves as a basis for workflow integration within the supply chain scenario. The data structure is used both for the intra- and inter-organizational production planning. In the case of intra-organizational planning tasks, the order data serve as dynamic input to the planning process[11]. The bill of materialist analyzed and, taking into account the operations another parameters like the lot size or due dates, partial orders are assigned to single production cells. [16].

MAS-based planning algorithms perform the planning process within the projects mentioned in the first chapter [12]. In the inter-organizational case, order data and estimated completion dates are exchanged between the partners to realize an integrated production planning. The DISPOWEB agents need information about the load situation and completion times from the particular intra-organizational agents. The ATT systems are triggered with the information derived from the planning processes and then begin to gather information of the monitored orders (e.g. milestones, event data) [15].

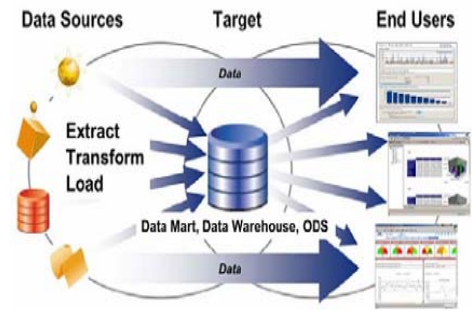

Fig.3 Database Structure

## 3. Integrated SCM Scheduling

To enable the accomplishment of production plan comprising the whole supply chain, we will use the concept of supply chain scheduling where each firm has to be represented by each firm as shown in fig.3.

➢ The production plan is initiated by occurrence of order (1).

➢ The DISPOWEB agent requests the KRASH agent for processing time & related cost for particular manufacturing process (2).

➢ After getting the task load information from KRASH [6] agent, the DISPOWEB agent begins the formulate the set of feasible scenarios (3).

➢ The best possible scenario with latest possible time is computed & submitted to DISPOWEB agent (4)

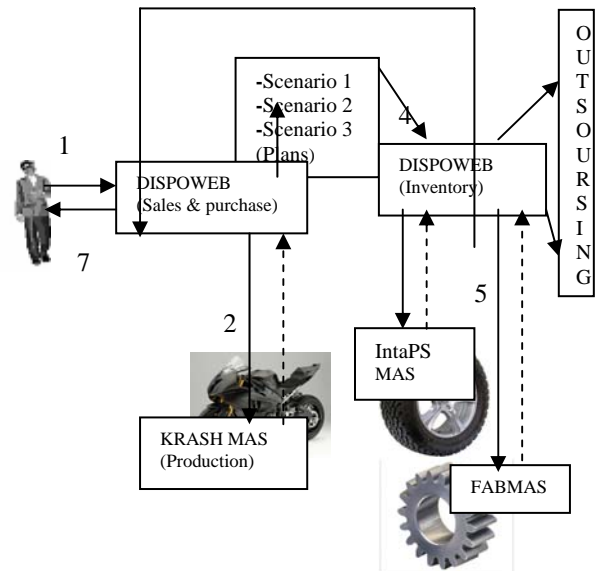

Fig 4.Supply Chain Scheduling with MAS

➢ This best scenario is further moved to other suppliers i.e. FABMAS [4] & IntaPS MAS for delivery of required input products through DISPOWEB agent (5).

➢ The DISPOWEB agent get the information from other agents that are not depend on other suppliers can fulfill the demand by submitting a supply vector consist of production cost & load situation for required sources to the final DISPOWEB agent (6).

➢ Finally the DISPOWEB agent computes the most profitable time for job execution is computed & dispatches the job by transmitting the selected delivery time to supplier agent, implying the closing of delivery contract between the parties. At last can contract to the consumer (7).

## 4. Achievements of MAS in Integrated SCM Architecture

The goal of the integrated SCM architecture is the co-ordination of the PPC (Production Planning and Control) activities throughout the whole logistics chain to realize a global monetary optimization. With the concept of database structure we can easily explore,extract,transform and deliver the data to anywhere at any time and frequency .With the help of various projects like DISPOWEB,FABMAS,KRASH,Inta-PS we create a successful production plan with the help of different multi agents that can interact with each other without any human intervention.

## 5. Conclusion

To compete in today's business environment, companies are required to integrate business activities into a global response strategy. To cope with sudden changes, companies need to conduct their business activities in a flexible manner. MAS is promised a solution framework for SCM because of the effective interaction between agents through an agent communication language. The integration of systems and technology also enables the reengineering of the business processes by leveraging technological capabilities that were previously unavailable. In this paper we show how MAS architecture interacts in the integrated SCM architecture with the help of various intelligent agents to highlight the above problem [8]. It addresses the architectural, the planning and the execution aspects of supply chain management. The goal of the evaluations is to prove the feasibility of the approach and gather first experiences and results [9]. In the next step, the system is evaluated using a real test case scenario that represents an industrial scope and the corresponding complexity. One instantiation of the database structure presented The supply chain that was chosen represents automobile manufacturer and its suppliers [5][10].

**Ritu Sindhu:** Ph.D Scholar, Banasthali University and WIT Gurgaon. Completed her B.Tech (CSE) from U.P.T.U, Lukhnow, M.Tech (CSE) from Banasthali University, Rajasthan.

**Abdul Wahid**: Presently working as a professor in Computer science department in AKG Engineering College Ghaziabad, India. Completed his MCA, M.Tech. and Ph.D. in Computer Science from Jamia Millia Islamia (Central University), Delhi.

**G.N.Purohit:** Dean, Banasthali University, Rajasthan.